\documentclass[sigconf]{acmart}
\usepackage{subcaption}
\usepackage{booktabs}
\usepackage{multirow}
\usepackage{graphicx}

\AtBeginDocument{%
  \providecommand\BibTeX{{%
    \normalfont B\kern-0.5em{\scshape i\kern-0.25em b}\kern-0.8em\TeX}}}

\copyrightyear{2024}
\acmYear{2024}
\setcopyright{rightsretained}
\acmConference[WWW '24 Companion]{Companion Proceedings of the ACM Web Conference 2024}{May 13--17, 2024}{Singapore, Singapore}
\acmBooktitle{Companion Proceedings of the ACM Web Conference 2024 (WWW '24 Companion), May 13--17, 2024, Singapore, Singapore}
\acmDOI{10.1145/3589335.3648309}
\acmISBN{979-8-4007-0172-6/24/05}

\makeatletter
\gdef\@copyrightpermission{
  \begin{minipage}{0.3\columnwidth}
   \href{https://creativecommons.org/licenses/by/4.0/}{\includegraphics[width=0.90\textwidth]{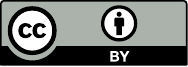}}
  \end{minipage}\hfill
  \begin{minipage}{0.7\columnwidth}
   \href{https://creativecommons.org/licenses/by/4.0/}{This work is licensed under a Creative Commons Attribution International 4.0 License.}
  \end{minipage}
  \vspace{5pt}
}
\makeatother

\begin{document}

%%
%% The "title" command has an optional parameter,
%% allowing the author to define a "short title" to be used in page headers.
\title{\modelname: Multi-Task Multi-Entity Embeddings for Pinterest Search}
%%
%% The "author" command and its associated commands are used to define
%% the authors and their affiliations.
%% Of note is the shared affiliation of the first two authors, and the
%% "authornote" and "authornotemark" commands
%% used to denote shared contribution to the research.

\author{Prabhat Agarwal}
\authornotemark[1]
\email{pagarwal@pinterest.com}
\orcid{0000-0002-3826-0858}
\affiliation{%
  \institution{Pinterest}
  \country{USA}
}

\author{Minhazul Islam SK}
\authornote{Both authors contributed equally to this work.}
\email{msk@pinterest.com}
\affiliation{%
  \institution{Pinterest}
  \country{USA}
}

\author{Nikil Pancha}
\email{npancha@pinterest.com}
\orcid{0000-0002-1755-7601}
\affiliation{%
  \institution{Pinterest}
  \country{USA}
}

\author{Kurchi Subhra Hazra}
\email{ksubhrahazra@pinterest.com}
\orcid{0009-0003-6929-2662}
\affiliation{%
  \institution{Pinterest}
  \country{USA}
}

\author{Jiajing Xu}
\email{jiajing@pinterest.com}
\orcid{0000-0002-4761-5171}
\affiliation{%
  \institution{Pinterest}
  \country{USA}
}
\author{Chuck Rosenberg}
\email{crosenberg@pinterest.com}
\orcid{0009-0003-9664-8644}
\affiliation{%
  \institution{Pinterest}
  \country{USA}
}

%%
%% By default, the full list of authors will be used in the page
%% headers. Often, this list is too long, and will overlap
%% other information printed in the page headers. This command allows
%% the author to define a more concise list
%% of authors' names for this purpose.
\renewcommand{\shortauthors}{Prabhat Agarwal et al.}
%% No italics

%%
%% The abstract is a short summary of the work to be presented in the
%% article.
\begin{abstract}
In this paper, we present \modelname, a versatile and scalable system for understanding search queries, pins, and products for Pinterest search. We jointly learn a unified query embedding coupled with pin and product embeddings, leading to an improvement of $>8\%$ relevance, $>7\%$ engagement, and $>5\%$ ads CTR in Pinterest's production search system. The main contributors to these gains are improved content understanding, better multi-task learning, and real-time serving. We enrich our entity representations using diverse text derived from image captions from a generative LLM, historical engagement, and user-curated boards. Our multitask learning setup produces a single search query embedding in the same space as pin and product embeddings and compatible with pre-existing pin and product embeddings.  We show the value of each feature through ablation studies, and show the effectiveness of a unified model compared to standalone counterparts. Finally, we share how these embeddings have been deployed across the Pinterest search stack, from retrieval to ranking, scaling to serve $300k$ requests per second at low latency. Our implementation of this work is available at this \href{https://github.com/pinterest/atg-research/tree/main/omnisearchsage}{link}\footnote{https://github.com/pinterest/atg-research/tree/main/omnisearchsage}.
\end{abstract}

%%
%% The code below is generated by the tool at http://dl.acm.org/ccs.cfm.
%% Please copy and paste the code instead of the example below.
%%
\begin{CCSXML}
<ccs2012>
   <concept>
       <concept_id>10002951.10003317.10003338</concept_id>
       <concept_desc>Information systems~Retrieval models and ranking</concept_desc>
       <concept_significance>500</concept_significance>
   </concept>
   <concept>
       <concept_id>10002951.10003260.10003261</concept_id>
       <concept_desc>Information systems~Web searching and information discovery</concept_desc>
       <concept_significance>500</concept_significance>
    </concept>
 </ccs2012>
\end{CCSXML}

\ccsdesc[500]{Information systems~Retrieval models and ranking}
\ccsdesc[500]{Information systems~Web searching and information discovery}

%%
%% Keywords. The author(s) should pick words that accurately describe
%% the work being presented. Separate the keywords with commas.
\keywords{Multi-Task Learning, Multimodal Embeddings, Representation Learning, Search Recommendation Systems}

%% A "teaser" image appears between the author and affiliation
%% information and the body of the document, and typically spans the
%% page.
% \begin{teaserfigure}
%   \includegraphics[width=\textwidth]{sampleteaser}
%   \caption{Seattle Mariners at Spring Training, 2010.}
%   \Description{Enjoying the baseball game from the third-base
%   seats. Ichiro Suzuki preparing to bat.}
%   \label{fig:teaser}
% \end{teaserfigure}

% \received{09 November 2023}
% \received[revised]{12 March 2009}
% \received[accepted]{5 June 2009}

\definecolor{lightcyan}{HTML}{E0FFFF}
\newcommand{\hlcyan}[1]{{\sethlcolor{lightcyan}\hl{#1}}}
\definecolor{lightgreen}{HTML}{DAF7A6}
\newcommand{\hlgreen}[1]{{\sethlcolor{lightgreen}\hl{#1}}}
\newcommand{\revision}[1]{{\color{black}#1}}
\newcommand{\modelname}{{\color{black}\textsc{OmniSearchSage}}}

%%
%% This command processes the author and affiliation and title
%% information and builds the first part of the formatted document.
\maketitle

\section{Introduction}

Pinterest’s mission is to bring everyone the inspiration to create a life they love.
Search is one of the key surfaces on Pinterest where users seek inspiration spanning a wide range of interests, such as decorating their homes, planning weddings, or keeping up with the latest trends in beauty and fashion. In order to enhance the search experience, modern search systems aim to incorporate various types of content such as web documents, news, shopping items, videos, and more. Similarly, Pinterest's search feed encompasses a diverse range of content, including pins, shopping items, video pins, and related queries.
To construct an inspiring feed for each of the more than $6$ billion searches per month on Pinterest we must uncover relevant content from billions of pins and products. We must also find relevant queries to help users refine their queries and navigate their search journey.

As an additional challenge, Pinterest search is global and multilingual with searchers using more than $45$ languages to find inspirational content.

Embeddings are useful building blocks in recommendation systems, especially search, where natural language understanding is key~\cite{fb2020embsearch,alibaba2020searchgraph,nigam2019semantic}.
Embeddings can power retrieval use cases via approximate nearest neighbor (ANN) search~\cite{malkov2018efficient,johnson2019billion}, enable detailed content and query understanding in ranking models without the overhead of processing raw data, and serve as a strong base to learn in low-data use-cases~\cite{ruder2019transfer-nlp}.
Despite their utility, embeddings come with their own challenges: if we learn a separate embedding for every use-case, there is an explosion of potentially expensive models that must be inferred on every request and used in downstream models.
This also may lead to suboptimal recommendation quality -- some use-cases may not have enough labels to learn an optimal representation.
In practice, it could entail additional maintenance costs and technical debt for upgrading to new versions of embeddings in certain applications, as some data may have been collected over the course of months or years.

Through rigorous offline experimentation, we show the impact of our key decisions in building embeddings for web-scale search at Pinterest:
\begin{itemize}
    \item Pin and product representations can be substantially enriched using diverse text derived from image captions from a generative LLM, historical engagement, and user-curated boards.
    \item A single query embedding can be used to retrieve queries, products, and Pins with nearly the same effectiveness as task-specific embeddings.
    \item A single query embedding can learn compatibility with multiple pre-existing embeddings \textit{and} learned entity embeddings, and perform well when compared across tasks.
    
\end{itemize}

\modelname{} has been deployed at Pinterest and is an integral component of the search stack.
It powers embedding-based retrieval for standard and product pins, queries and ads.
It is also one of the most important feature in multi-stage ranking models and various query classification models.
These gains all arise despite the existence of other features enabling pin and product understanding, which highlights the importance optimizing embeddings end-to-end for search.

\section{Related Work}
Our work to build multi-task multi-entity embeddings for search draws upon broad areas of work.
Our representation of pins and products extends existing work on multi-modal learning and two tower models for search retrieval. 
These have been extensively applied in the context of search and recommendation systems as an efficient way to retrieve results not purely related to the search query based on text.
In \modelname, we demonstrate that the embeddings generated by these models can also serve as features in ranking and relevance models.
Additionally, we offer a brief examination of specific embeddings within the Pinterest ecosystem.

\subsection{Model-based Search Retrieval}
Historically, search systems have been powered by two stages: token-based matching, or candidate generation, and then scoring with a complex model.
These have drawbacks, especially when users make complex queries or content is not primarily textual.
This has led to the exploration of two tower models, which encode a query into a single embedding or a small set of embeddings, and then use those to retrieve relevant documents with approximate or exact nearest neighbor search~\cite{alibaba2020searchgraph, fb2020embsearch, amzn2022graphmultilingual, zhang2022uni, li2021embedding, lu2020twinbert, magnani2022semantic}.

Two natural topics in learning embeddings for search are document representation, and query representation.
Depending on the learning objective, this query representation could be personalized, or it could be a pure text embedding model.
Many architectures for query embeddings in industry have been proposed based on simple CNNs~\cite{msft2013dssm}, bag of words models~\cite{fb2020embsearch,nigam2019semantic}, transformers~\cite{fb2021que2search}, and more, but they share a basic structure involving query understanding and sometimes context understanding.
Document representation is also a major challenge.
The text associated directly with an item is popular as a key feature, but depending on the task, other sources have been found to provide great value, including queries where other users have engaged with a given item~\cite{alibaba2020searchgraph,amzn2022graphmultilingual, nogueira2019doc2query} and image content embeddings~\cite{fb2021que2search}.

\subsection{Multi-task, multi-modal, and multi-entity embeddings}
The area of learning embeddings isn't exclusive to the realm of recommendation systems and has been studied extensively~\cite{radford2021learning, reimers2019sentence, cer2018universal, conneau2017supervised}.
Multi-task learning is a technique commonly utilized in ranking models to optimize for multiple objectives concurrently, aiming for enhanced performance or more efficient information sharing~\cite{youtube2019multitask,tang2020ple}.
A less frequently encountered approach involves the joint learning of embeddings for more than two entities. Though this methodology is sometimes implemented in graph learning scenarios, it can also be perceived as an extension of multi-task learning~\cite{hetgnn2019}.

Multi-modal embeddings are of substantial interest in the industry since the majority of web content is multi-modal, typically including at both text and images~\cite{fb2022commercemm, fb2021que2search, li2021embedding}.
One can take embeddings or raw data from each modality as inputs, and merge them at any stage of the model.
The methodology typically involves utilizing embeddings or raw data from each mode as inputs, which are then merge at different stages in the model.
Early-stage fusion can pose computational hurdles; therefore, in cases where performance is indifferent, utilizing embeddings instead of raw data is generally the preferred course of action~\cite{fb2022commercemm}.

\subsection{Embeddings at Pinterest}
PinSage~\cite{ying2018graph} is a scalable GNN-based embedding representing pins. It is based on the GraphSage GCN algorithm~\cite{hamilton2017graphsage}, sampling neighborhoods with personalized PageRank to augment pin understanding, instead of simple heuristics like $n$-hop neighbors.
It aggregates some basic visual~\cite{beal2022billion} and text information into a single dense representation, and is a critical feature in many models.

To represent products, we have an embedding, ItemSage~\cite{baltescu2022itemsage}, which aggregates raw data about products, including metadata from product pages, and potentially many images of the product.
ItemSage is trained for compatibility with PinSage, and the search query embedding preceding \modelname{}, meaning that the distance between ItemSage and these two embeddings can be used for retrieving or ranking content~\cite{searchsage_2021}.
\section{Method}

\subsection{Problem Formulation}
In order to enhance the search experience, modern search systems aim to incorporate various types of content such as web documents, news, shopping items, videos, and more. Similarly, Pinterest's search feed encompasses a diverse range of content, including pins, shopping items, video pins, and related queries. Training separate query embedding models for each content type and its representation proves to be resource-intensive and inefficient. To address this issue, we introduce \modelname, which offers a unified query embedding model that jointly trains query embeddings for query-query, query-pin, and query-product retrieval and ranking. 

Another requirement in production systems is compatibility with existing embeddings, which is essential for purposes such as cost-efficiency and simplified migration. Hence we also train the query embeddings to be compatible with the corresponding pre-existing embeddings for the entities. As a side effect, we also get compatibility with some embeddings due to the triangle inequality property inherent to cosine similarity.

\subsection{Enriching Entity Representations}\label{sec:text_enrichment}

On Pinterest, each pin or product is associated with an image and title, along with an optional text (known as description) and link. Beyond these typical attributes, products may carry additional metadata, such as brand information, color description, and more. 
Document expansion techniques has been empirically demonstrated to significantly enhance the performance of not just token-based, but also embedding-based search retrieval systems~\cite{efron2012improving, nogueira2019doc2query, tao2006language, pickens2010reverted, nogueira2019document}. Hence, in \modelname, we enrich our entity representations using diverse text derived from image captions from a generative LLM, historical engagement, and user-curated boards as described below. In the dataset, $71\%$ of pins and products feature a title or description, $91\%$ include non-empty board titles, and $65\%$ contain non-empty engaged queries. Synthetic GenAI captions are generated for all pins and products, ensuring full coverage. Section~\ref{sec:text_ablations} discusses the importance of each of these enrichment.

\subsubsection{Synthetic GenAI Captions}
On our platform, a substantial volume of pins (about $30\%$) lack associated titles or descriptions, or possess noisy and/or irrelevant title or description.
We address this issue by employing an off-the-shelf image captioning model, BLIP~\cite{li2022blip}, to generate synthetic descriptions for these images. 

To assess the quality of these synthetically generated descriptions, we enlisted human evaluators to judge their relevance and quality. For a robust assessment, three distinct ratings were collected for each image within a sample of $10k$ images, curated uniformly across various broad pin categories. The results indicated that an overwhelming $87.84\%$ of the generated descriptions were both relevant and of high quality, while a meager $1.16\%$ were deemed irrelevant and of poor quality. 

These synthetically generated descriptions serve as an added feature in our model, enriching the diversity of data associated with each entity. Despite not being directly visible to the users, their addition significantly contributes to a deeper understanding of the pins' content.

\subsubsection{Board Titles}
On Pinterest, users explore and save pins to their personal collections, referred to as boards. Each board carries an associated title, reflecting the topic or theme of the collection. Most often, these user-crafted boards are meticulously organized, each focusing on a distinct theme or purpose. A user might, for instance, create discrete boards for \textit{``Social Media Marketing"} and \textit{``Graphic Design'\"}. Consequently, these board titles provide valuable, user-generated descriptors for the pins within the respective boards.

We exploit this user-curated information by accumulating the titles of all boards each pin has been saved to. We limit our selection to a maximum of $10$ unique board titles for each pin/product, systematically eliminating any potentially noisy or redundant titles as described next. First, each title is assigned a score influenced by two factors: its frequency of occurrence and the prevalence of its comprising words. Following this, titles are then ranked based on a hierarchy of their score (ascending), word count (descending), and character length (descending). The resulting top $10$ board titles are subsequently incorporated as a feature in our model. This process eliminates any potentially noisy or redundant titles from the feature.

\subsubsection{Engaged Queries}

When multiple users interact with a specific pin or product for a certain query within a search feed, it signifies that pin's relevance to that query. We can use these queries to expand our understanding of the pin/product. For every pin, we generate a list of queries that have attracted user engagements, along with the counts and types of such engagements. This list of queries is then sorted using a function based on the count for each type of engagement. We use the top $20$ queries from these sorted lists as a feature in our model.

Through experimentation with diverse time-windows of query logs for feature creation, we discovered that larger windows yield superior performance. Consequently, we have opted for a two-year window for feature calculation. However, the complexity of computing this from scratch every time presents a challenge. To mitigate this, we deploy an incremental approach. Every $n$ days, we examine new query logs, create a list of queries for every pin, and then blend it with the previously existing top $20$ queries, thereby updating the latest value of the feature.

\subsection{Entity Features}

The features we incorporate include PinSage~\cite{ying2018graph} and unified image embeddings~\cite{beal2022billion} to capture the essence of each pin. Additionally, for product pins, we use ItemSage~\cite{baltescu2022itemsage} given its capability in effectively representing product-related pins. Text-based features such as the title and description of each pin are also integral to our feature set. Furthermore, we augment the text associated with each pin with the inclusion of synthetic captions, board titles, and engagement queries as outlined earlier. By integrating all these features, we attain a comprehensive and multi-dimensional representation of each pin, hence facilitating enhanced learning of representations.

\begin{figure}
    \centering
    \includegraphics[width=0.9\linewidth, scale=0.7, trim={30, 30, 30, 30}, clip]{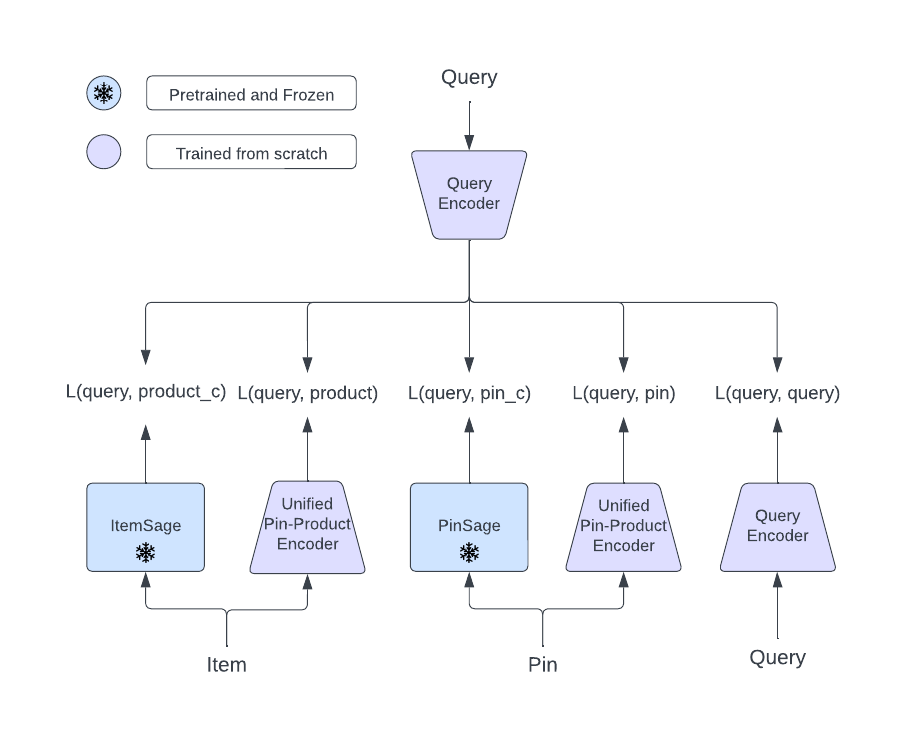}
    \caption{Diagrammatic Representation of \modelname{}'s Multi-Entity, Multi-Task Architecture.}
    \label{fig:loss_multitask}
\end{figure}

\subsection{Encoders}
In our work, we consider $3$ entity types, namely, pin, product and query. Our model consists of an encoder for query, a unified learned encoder for both pin and product, and dedicated compatibility encoders for pin and product, respectively.

\subsubsection{Query Encoder}

\begin{figure}
    \centering
    \includegraphics[width=0.9\linewidth, scale=0.3, trim={5 35 0 10}]{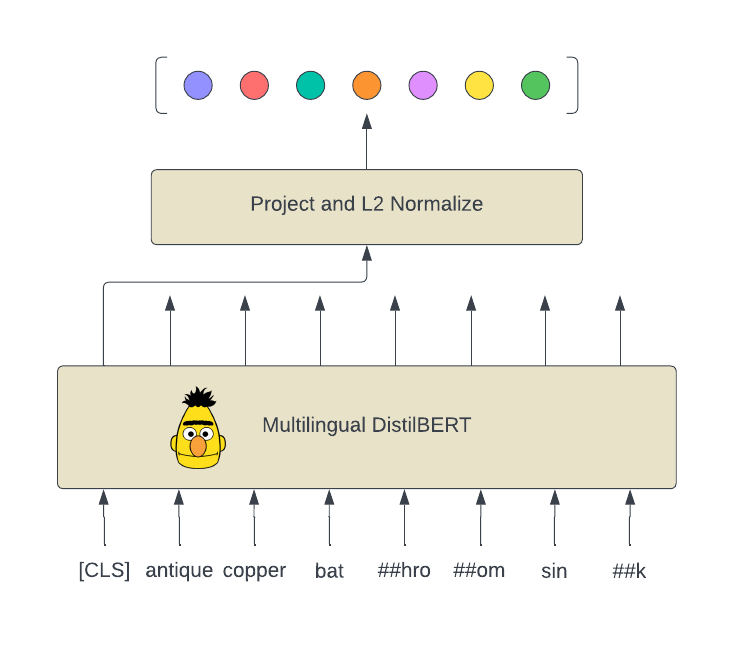}
    \caption{Overview of the query encoder architecture. The encoder takes the output from the last layer associated with the `CLS' token, projects it onto a 256-dimensional vector space, and finally L2-normalizes the output to generate the final embedding.}
    \label{fig:query_encoder}
\end{figure}
The query encoder in our model (depicted in Figure~\ref{fig:query_encoder}) is based on a multilingual version of the DistilBERT (\textit{distilbert-base-multilingual-cased}\footnote{\url{https://huggingface.co/distilbert-base-multilingual-cased}})~\cite{Sanh2019DistilBERTAD}. This choice facilitates efficient handling of queries across a variety of languages.
The encoder utilizes the output from the last layer corresponding to the $CLS$ token and thereafter projects it to a $256$-dimensional vector space. Post projection, we apply a $L2$ normalization on the $256$-dimensional vectors to obtain the final embedding. This normalization greatly simplifies the calculation of cosine-distance in downstream applications, allowing for a straightforward dot product operation.

\subsubsection{Unified Pin and Product Encoder}\label{sec:unified-model}
\begin{figure}
    \centering
    \includegraphics[width=0.9\linewidth, scale=0.9, trim={5 35 0 10}]{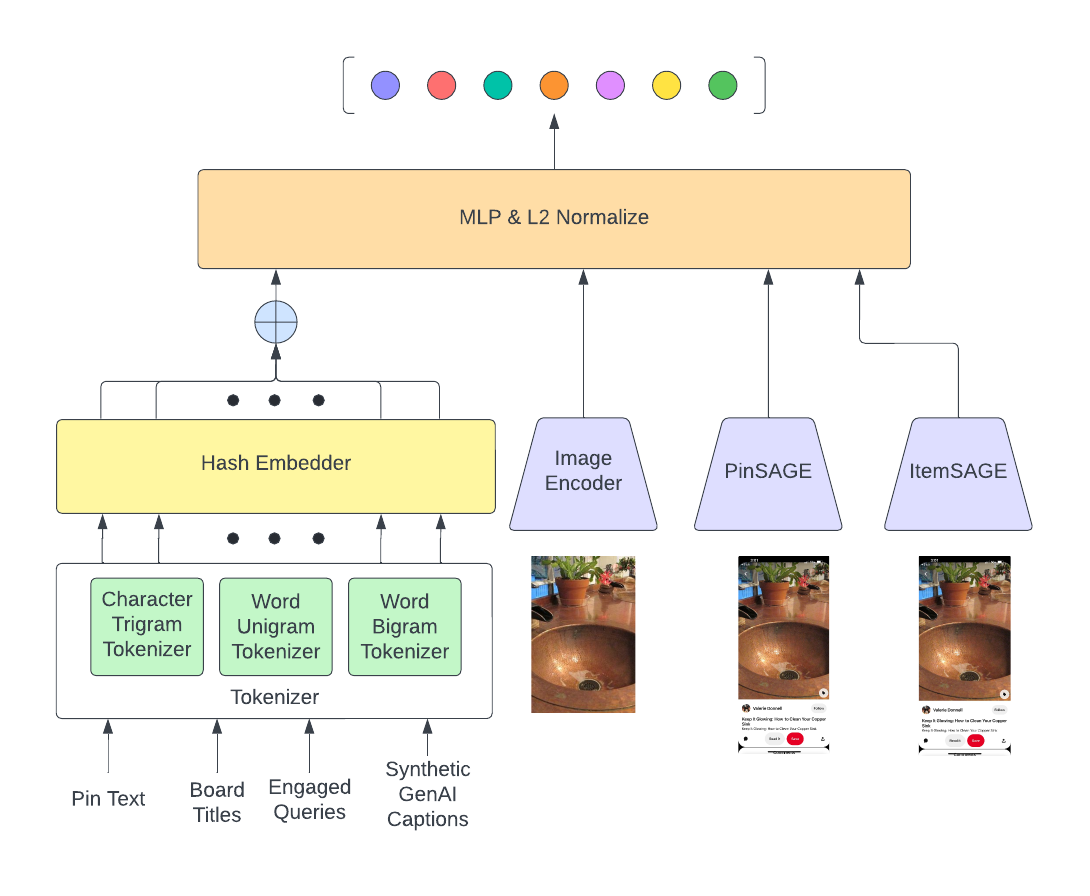}
    \caption{Schematic of the unified encoder model for pins and products, illustrating the use of three different tokenizers, a hash embedding table, and an MLP layer for combining text embeddings with other continuous features.}
    \label{fig:unified_pin_item_encoder}
\end{figure}

In our model, we utilize a single unified encoder for both pins and products (depicted in Figure~\ref{fig:unified_pin_item_encoder}), and this encoder is jointly trained with the query embeddings. Designed to process both textual features and continuous features, it plays a crucial role in learning the respective embeddings of pins and products. In cases where certain features are defined for one entity but not the other, we substitute them with zero, ensuring a consistent data input. 

As detailed in section~\ref{sec:loss}, we utilize in-batch negatives to train our model. Prior research~\cite{NEURIPS2020_d89a66c7, lee-etal-2019-latent, giorgi-etal-2021-declutr, radford2021learning} has empirically demonstrated that larger batches with a substantial number of negatives help in learning better representations. Therefore, to accommodate a larger batch size in the GPU memory, we employ a simple pin encoder model. The following encoder design has been determined through numerous ablation studies. These studies have allowed us to select the most effective configuration for each of the components, while still considering the importance of both training and serving efficiencies.

The encoder uses three distinct tokenizers to process the textual features associated with a pin~\cite{baltescu2022itemsage, nigam2019semantic, huang2013learning}. These include (i) a word unigram tokenizer that uses a vocabulary encompassing the $200k$ most frequent word unigrams, (ii) a word bigram tokenizer that makes use of a vocabulary comprising the $1M$ most frequent word bigrams, and (iii) a character trigram tokenizer that utilizes a vocabulary of $64k$ character trigrams. The tokens are mapped to their respective IDs in the vocabulary $\mathcal{V}$ which constitute all three tokenizers. Any token that falls out of this combined vocabulary gets discarded. The use of these combined tokenizers effectively helps in capturing the semantics of various texts associated with a pin/product.

For token embedding learning, we use a $2$-hash hash embedding table of size $100,000$~\cite{tito2017hash, baltescu2022itemsage}. Each identified token's ID $i$ is hashed into two places within the embedding table using hash functions $h_1(i)$ and $h_2(i)$. The ultimate embedding of a token with ID $i$ is a weighted interpolation of the two locations: $W_{1i} h_1(i) + W_{2i} h_2(i)$, where $W_1$ and $W_2$ are learned weight vectors of size $|\mathcal{V}|$ each.

The sum of all token embeddings and the embedding features are concatenated and fed into a 3-layer MLP, with layer sizes of $1024, 1024, 256$. Following this, the output of the MLP layer undergoes L2-normalization just like the query embedding.

\subsubsection{Compatibility Encoders}
In our model, we employ two discrete compatibility encoders individually dedicated to pins and products. These encoders leverages the  pre-existing pin and product embeddings, represented by PinSage for pins and ItemSage for products. This allows the model to adeptly learn query embeddings that align effectively with PinSage and ItemSage embeddings.

\subsection{Multi-Task Sampled Softmax Loss} \label{sec:loss}

Taking inspiration from Itemsage~\cite{baltescu2022itemsage}, the problem of learning query and entity embeddings is treated as an extreme classification problem, with the aim of predicting entities relevant to a given query~\cite{covington2016deep}. We employ the sampled softmax loss with logQ correction~\cite{yi2019sampling} to train our model.

We use multitasking to jointly train entity embeddings and train the query embeddings to be compatible with existing entity embeddings.

Formally, we define a task $T \in \mathcal{T}$ as a tuple of a dataset of query-entity pairs ($\mathcal{D} = \{ (x, y)_i \}$) and an entity encoder $\mathcal{E}$.
\vspace{-3pt}
$$T \triangleq \{ \mathcal{D}, \mathcal{E} \}.$$
For a batch of data, $\mathcal{B} = \{(x, y)_i\} \subset \mathcal{D}$, for task $T \in \mathcal{T}$, the aim is to learn query embedding $q_{x_i}$ and entity embedding $p_{y_i} = \mathcal{E}(y_i)$ such that the cosine similarity of the embeddings $q_{x_i} \cdot p_{y_i}$ is maximized. This is achieved by minimizing the softmax loss:
\begin{equation}
L_T = - \frac{1}{|\mathcal{B}|} \sum_{i=1}^{|\mathcal{B}|} \log \frac{\exp(q_{x_i} \cdot p_{y_i})}{\sum_{y \in C} \exp(q_{x_i} \cdot p_{y})},
\end{equation}
where $\mathcal{C}$ is the catalog of all entities of the same type as $y_i$.
To ensure problem tractability, the normalization term in the denominator is approximated using a sample of the catalog $C$. We use (i) positives in the batch, $BN = \{y_i | (x_i, y_i) \in \mathcal{B}\}$, and (ii) a random sample of the catalog, $C'$. To rectify any bias that might have been introduced through sampling, we utilize the logQ correction technique. This method operates by deducting the sampling probability of the negative, represented as $\log Q(y|x_i)$, from the existing logits. This is crucial to ensure that popular entities aren't disproportionately penalized.

\begin{align}
    L_{T} &= L^{S_{bn}}_{T} + L^{S_{rn}}_{T} \\
    L^{S_{bn}}_{T} &= - \frac{1}{|\mathcal{B}|} \sum_{i=1}^{|\mathcal{B}|} \log \frac{\exp(q_{x_i} \cdot p_{y_i} -\log Q(y_i|x_i))}{\sum_{z \in BN} \exp(q_{x_i} \cdot p_{z} -\log Q(z|x_i))} \\
     L^{S_{rn}}_{T} &= - \frac{1}{|\mathcal{B}|} \sum_{i=1}^{|\mathcal{B}|} \log \frac{\exp(q_{x_i} \cdot p_{y_i} -\log Q(y_i|x_i))}{\sum_{y \in C^{'}} \exp(q_{x_i} \cdot p_{y} -\log Q(y|x_i))} \\
    &=  - \frac{1}{|\mathcal{B}|} \sum_{i=1}^{|\mathcal{B}|} \log \frac{\exp(q_{x_i} \cdot p_{y_i} -\log Q(y_i|x_i))}{\sum_{y \in C^{'}} \exp(q_{x_i} \cdot p_{y} -\log Q_n(y))},  \\
    &\text{since $y$ is sampled independently} \nonumber
\end{align}

The total loss is defined as the sum of all individual task losses, 
\vspace{-3pt}
\begin{equation}
     L = \sum_{T \in \mathcal{T}} L_T.
\end{equation}
We mix together different tasks together in one batch and control the influence of each task on the model through this composition.
To increase training efficiency, we share the pairs in the batch across all tasks with the same dataset.

\subsection{Model Serving}

\modelname{} query embeddings are integral to numerous applications in the search stack, which necessitates us to maintain a strict latency budget. 
For real-time inference with minimized latency, our query encoder is served on GPUs by our in-house C++-based machine learning model server, the Scorpion Model Server (SMS). Factoring in that query distribution complies with Zipf's law, we have instituted a cache-based system to curb costs and shorten response times.   The query embedding server first verifies if a query is cached before resorting to the query inference server should it be absent from the cache. After testing various Cache Time-To-Live (TTL) periods, a TTL of $30$ days was established as optimal.
The system is equipped for handling $300k$ requests per second, maintaining a median (p50) latency of just $3$ms, and $90$ percentile (p90) latency of $20$ms. The implementation of this cache-based system efficiently reduces the load on the inference server to approximately $500$ QPS, leading to substantial cost and latency reductions.

The pin and product embeddings are derived offline on a daily basis through batch inference on GPUs and are subsequently published to our signal store for consumption.

\section{Experiments}

\subsection{Dataset}
Our dataset is primarily constructed by extracting unique query-entity pairs from one year of search query logs. We consider various forms of engagement on the platform when extracting these pairs, including `saves' (when a user saves a pin to a board) and `long clicks' (instances where users browse the linked page for more than $10$ seconds before returning to Pinterest). For products, we enrich our dataset by incorporating offsite actions as well. Thus, we also include anonymized pairs tied to significant actions like `add to cart' and `checkout'. A common challenge in recommendation systems is the popularity bias, where certain pins are overrepresented due to their high appeal. To counteract this bias, we impose a limit on the number of times the same pin can be paired. This limit is capped at $50$ pairs for pins and is extended to $200$ pairs for products (since products have lower volume and engagement). By adopting this strategy, we ensure our dataset is robust and truly representative of the user's activity on the platform.

Our model training is further extended to encompass query-query pairs. On Pinterest, users are presented with similar query suggestions, and engagements with these recommendations are recorded in the search logs. We leverage these records, extracting such pairs from an entire year's logs, thus enriching our training dataset. 

A detailed breakdown of the positive labels in the dataset is provided in Table~\ref{tab:data_stats}.
\begin{table}
    \centering
    \begin{tabular}{cccc}
    \toprule
         Pair & Source & Actions& Size\\
         \midrule
         Query-Pin & Query Logs & repin, longclick& 1.5B\\
         Query-Product & Query Logs & repin, longclick & 136M\\
         Query-Product & Offsite logs & add-to-cart, checkout & 2.5M \\
         Query-Query & Query Logs & click & 195M \\
         \bottomrule
    \end{tabular}
    \caption{Summary of the different training datasets.}
    \label{tab:data_stats}
\end{table}

\subsection{Offline Evaluation Metrics}
Our evaluation of the model encompasses both user engagement data and human-labeled relevance data.

Relevance gets measured using human-labeled pairs of queries and pins, sampled from production traffic from four distinct countries: US, UK, France, and Germany. This strategy serves to assess the model's performance in handling multiple languages and cultural contexts.

Evaluation of user engagement considers a selected $7$-day period. We ensure no data leakage—possible due to the inclusion of engagement features such as engaged queries—by maintaining a $15$-day separation between the end of the training dataset and the beginning of the evaluation phase. We sample $80k$ pairs from the defined evaluation duration to represent repins and long clicks for both pins and products. Another $80k$ pairs, corresponding to clicks for queries, are also included for comprehensive performance evaluation.

The primary metric we used for evaluation is named `Recall@10'. This metric denotes the likelihood of the occurrence of the engaged entity within the top 10 entities when these entities are sorted in descending order based on their similarity to the query.

Consider a dataset $D = {(q_i, e_i)}_{i=1}^{n}$, where each $(q_i, e_i)$ denotes a query-engaged entity pair, and also consider a random corpus $C$ with $m$ entities. The Recall@10 metric can then be defined as the average over all queries of the indicator function $\mathbf{1}$, where $\mathbf{1}$ equals 1 if the engaged entity $e_i$ is amongst the top $10$ entities in $C$ when ranked by their dot product with the query $q_i$.
$$
\text{Recall@}10= \frac{1}{|D|}\sum_{i=1}^{|D|} \mathbf{1} [(\sum_{y \in C} x_i \cdot y > x_i \cdot y_i) > 10]
$$

For every pin, query, and product, we employ a uniformly distributed random sample of $m=1.5M$ entities from our corpus.

\subsection{Offline Results}

In this section, we provide a comprehensive comparison between our proposed model, \modelname, and the existing baselines, which helps showcase its performance enhancements. Subsequently, we undertake an in-depth exploration of key influential aspects such as the significance of text enrichments, the pros and cons of adopting multitasking approaches, and the operational efficacy of compatibility encoders in the context of our model.

\subsubsection{Comparison with Baselines}
In this study, the existing version of SearchSage~\cite{searchsage_2021} serves as our comparison baseline. It operates using fixed PinSage and ItemSage embeddings for pins and products, respectively. For \modelname{}, we utilize the query encoder to derive query embeddings and the unified pin and product encoder to generate pin and product embeddings.

\begin{table}[t]
\begin{tabular}{lccc}
\toprule
Metric & SearchSage & \modelname{} & Gain \\
\midrule
\multicolumn{4}{c}{Pin} \\
Save & $0.39$ & $0.65$ & $+67\%$ \\
Long-Click & $0.45$ & $0.73$ & $+62\%$ \\
Relevance (US) & $0.25$ & $0.45$ & $+80\%$ \\
Relevance (UK) & $0.29$ & $0.51$ & $+76\%$ \\
Relevance (FR) & $0.23$ & $0.43$ & $+87\%$ \\
Relevance (DE) & $0.28$ & $0.46$ & $+64\%$ \\
\midrule
\multicolumn{4}{c}{Product} \\
Save & $0.57$ & $0.73$ & $+28\%$ \\
Long-Click & $0.58$ & $0.73$ & $+26\%$ \\
\midrule
\multicolumn{4}{c}{Query} \\
Click & $0.54$ & $0.78$ & $+44\%$ \\
\bottomrule
\end{tabular}
\caption{Comparative analysis of \modelname{} and the baseline SearchSage across various tasks - Pin, Product, and Query.}
\label{table:baseline_results}
\end{table}

In Table~\ref{table:baseline_results}, comparisons are drawn between \modelname{} and SearchSage, with both models being trained and evaluated on the same dataset. It is important to highlight that the baseline model, SearchSage, does not involve query-query pairs for training purposes.

On the pin dataset, \modelname{} shows a significant gain, between $60\%$ and $90\%$, over SearchSage across all metrics. Recall is relatively consistent across different countries, reflecting the multilingual robustness of \modelname{}.

Analysis of the product dataset reveals that \modelname{} outperforms the baseline model by about $27\%$ in predicting product engagement. This increment is less prominent as compared to the pins dataset, mainly because ItemSage, upon which this comparison is based, has already undergone training on search tasks. Nevertheless, the observed improvement shows the positive impact of incorporating new features as well as the benefit of multi-tasking.

Interestingly, SearchSage is able to predict related query clicks substantially better than random despite not being trained on this task. However, when we directly optimize for this objective in \modelname{}, we see a substantial $+44\%$ improvement. We show this improvement can be attributed to both training on related queries, and multi-task learning in Section \ref{results:multitask}.

\subsubsection{Importance of content enrichment}\label{sec:text_ablations}

In this section, we delve into an analysis of the importance of various text enhancements described in Section~\ref{sec:text_enrichment}. To maintain brevity, the evaluation focuses solely on the metrics related to the query-pin task.

Our first direction of investigation centers around the impact of integrating synthetic captions for pins that lack both a title and description.
For this purpose, we extracted pairs from the evaluation dataset in which the engaged pin was missing a title or a description. This resulted in a narrowed evaluation dataset of $24k$ pairs. The model's performance, initially based on solely continuous features and native text, was then compared to a model additionally enriched with captions.

Table~\ref{tab:caption_perf} presents the results of this comparison. When synthetic captions were added, both `save' and `long-click' metrics saw substantial improvements — approximately $+30\%$ and $+26\%$ respectively. However, the relevance metric remained unchanged.

This suggests that adding synthetic captions can significantly enhance the model's performance for certain metrics when representing pins that lack a title and description.

\begin{table}
\centering
\begin{tabular}{@{}lccc@{}}
\toprule
\textbf{}     & save & long-click & relevance \\ \midrule
No captions   & 0.51          & 0.60                & 0.36               \\
With captions & 0.66          & 0.76                & 0.36               \\ \midrule
Improvement   & +30.43\%      & +25.58\%            & 0\%        \\ \bottomrule
\end{tabular}
\caption{Comparative assessment displaying the influence of Synthetic GenAI Captions on pins lacking titles and descriptions.}
\label{tab:caption_perf}
\end{table}
\begin{table}[b]
\centering
\resizebox{\columnwidth}{!}{%
\begin{tabular}{@{}lccc@{}}
\toprule
\multicolumn{1}{c}{\textbf{}}       & save & long-click & relevance \\ \midrule
Continuous Features Only & 0.43          & 0.53           & 0.30               \\
Adding Title, Description and Synthetic GenAI Captions                    & 0.52 (+21\%)  & 0.63 (+19\%)   & 0.39 (+30\%)       \\
Adding Board Titles                 & 0.61 (+17\%)  & 0.68 (+8\%)    & 0.44 (+13\%)       \\
Adding Engaged Queries              & 0.65 (+7\%)   & 0.73 (+7\%)    & 0.46 (+5\%)        \\ \bottomrule
\end{tabular}%
}
\caption{Impact of adding different text enrichments on the model's performance. Each percentage increase is relative to the previous row, displaying the additional improvement from each additional feature. }
\label{table:text_enrichment_ablations}
\end{table}
Table~\ref{table:text_enrichment_ablations} illustrates the impact of adding different text enrichments on the model's performance. Each percentage increase is relative to the previous row, displaying the additional improvement from each additional feature.

Our baseline model utilizes only continuous features for training and its performance values are reflected in the first row. Upon adding `Title', `Description', and `Synthetic GenAI Captions' to the baseline model, we notice a robust improvement across all metrics. There is a $20\%$ improvement in the engagement datasets, while the relevance metric improves by a notable $30\%$, demonstrating the substantial impact of these text features.

The model enhancement continues with adding board titles to the feature set, leading to a further increase of $8-15\%$ in different metrics. This affirms the relevance of board titles in improving predictive accuracy. 

Finally, we incorporated engaged queries feature into the model, resulting in a consistent, albeit smaller growth across all three metrics. Although the incremental relative gain appears smaller, it still constitutes a significant improvement when compared to the baseline model.
In summary, each text enrichment feature contributes significantly to improving model performance as seen by the increment in metrics compared to their immediate preceding state. 
\vspace{-5pt}
\subsubsection{Effect of multi-tasking}\label{results:multitask}
In Table~\ref{table:multitasking}, we present a comparative analysis between models trained independently for each task (pin, product, and query) and our consolidated multitask model. For this comparison, both the independent and multitask models were trained under equivalent conditions - with matching batch sizes, computational power, and iterations. The datasets used for both training and evaluation were also identical, with the sole difference that the individual models were trained on their respective subset of pairs from the dataset. This systematic approach ensures the fair and accurate assessment of the performance of the multitask model in relation to the independent task models.

On the pin task, we see slight degradation in quality from multi-task learning, but, on product and query tasks, results are neutral to positive. This aligns with general notions about multi-task learning: low-data tasks are unlikely to see regressions from multi-task learning, while the pin task using $1.5B$ pairs sees a very slight drop in performance. Despite this drop, the simplification benefits of multi-task learning outweigh the metric loss.

\begin{table}
\resizebox{\columnwidth}{!}{
\begin{tabular}{@{}lllllll@{}}
\toprule
\multicolumn{3}{l}{Dataset}                                  & Pin Only & Product only & Query Only & OmniSearchSage \\ \midrule
pin                      & \multicolumn{2}{l}{save}          & 0.68     & -         & -          & 0.65           \\
                         & \multicolumn{2}{l}{long-click}    & 0.75     & -         & -          & 0.73           \\
                         & \multicolumn{2}{l}{avg relevance} & 0.45     & -         & -          & 0.46           \\
                         \cmidrule(lr){1-7}
\multirow{2}{*}{product} & \multicolumn{2}{l}{save}          & -        & 0.73     & -          & 0.73           \\
                         & \multicolumn{2}{l}{long-click}    & -        & 0.73      & -          & 0.73           \\ \cmidrule(lr){1-7}
query                    & \multicolumn{2}{l}{click}         & -        & -         & 0.73       & 0.78           \\ \bottomrule
\end{tabular}}
\caption{Comparative analysis illustrating the contrasts between our unified multi-task model and models trained individually for each task - pin, product, and query.}
\label{table:multitasking}
\end{table}

\subsubsection{Effect of compatibility encoders}
We examine the influence of incorporating compatibility encoders on the effectiveness of the learned pin/product embeddings. We train a model that comprises only the query and unified pin and product encoder. Subsequently, this model is compared with another model that fully incorporates all the encoders. Interestingly, there is almost no noticeable degradation in the metrics of the learned encoder, thereby essentially achieving seamless compatibility of the query embedding with pre-existing embeddings at no substantial cost.

Furthermore, as demonstrated in Table~\ref{tab:compatibility_encoder}, the performance of the compatibility encoders in the \modelname{} model is either on par with or surpasses that of the SearchSage model, which is trained utilising only compatibility encoders.

\begin{table}[htb]
% \resizebox{\columnwidth}{!}{%
\begin{tabular}{@{}llcc@{}}
\toprule
\multicolumn{2}{l}{Dataset}                         &        SearchSage & \modelname{} \\ \midrule
pin                      & {save}          & 0.39        & 0.39           \\
                         & {long-click}    & 0.45     & 0.43           \\
                         & {avg relevance} & 0.26     & 0.26           \\
                         \cmidrule(lr){1-4}
\multirow{2}{*}{product} & {save}          & 0.57     & 0.57           \\
                         & {long-click}    &  0.58      & 0.57           \\  \bottomrule
\end{tabular}%
% }
\caption{Comparison of co-trained compatibility encoders with independently trained compatibility encoders.}
\label{tab:compatibility_encoder}
\end{table}

\section{Applications in Pinterest Search}

\begin{figure}[t]
    \centering
    \includegraphics[width=\linewidth, scale=0.6, trim={0, 20, 0, 0}, clip]{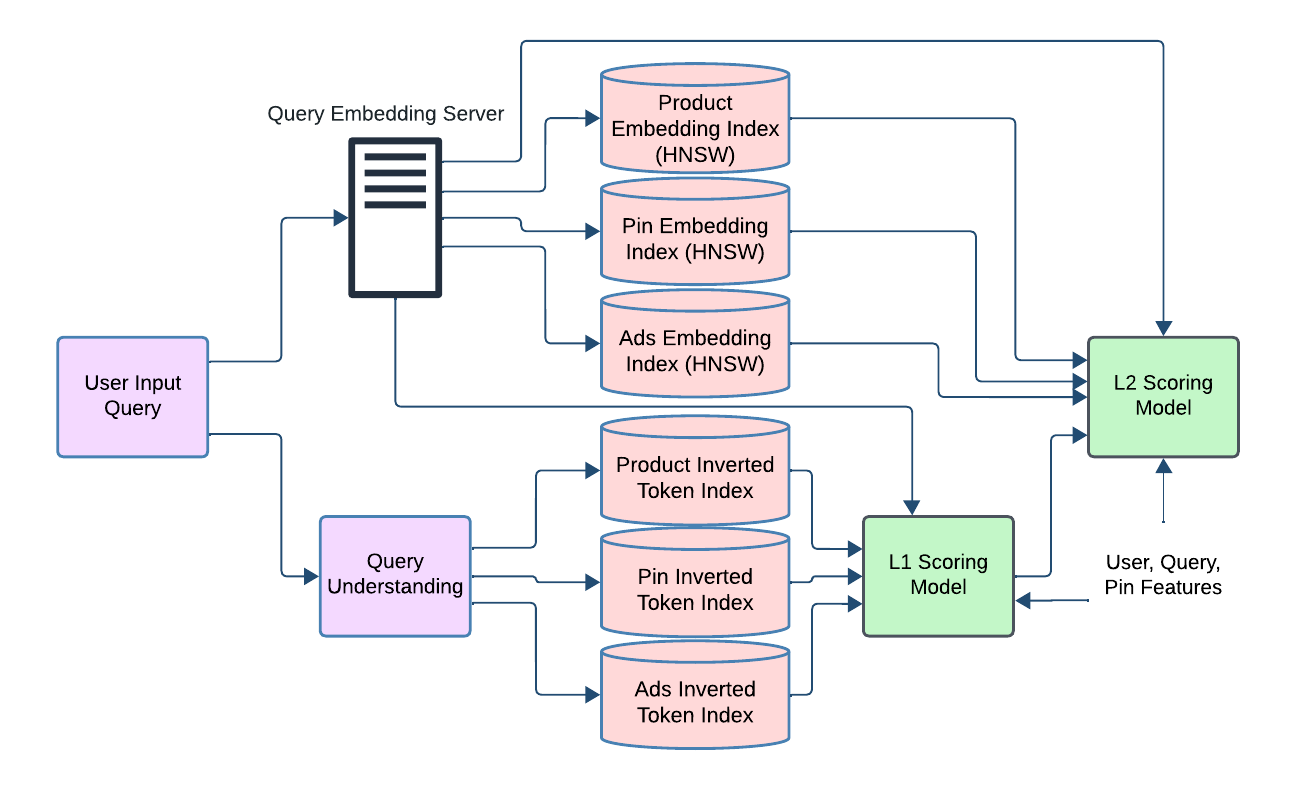}
    \caption{A simplified depiction of the search retrieval and ranking stack at Pinterest highlighting the integration points for \modelname{} embeddings.}
    \label{fig:stack}
\end{figure}

\modelname{} embeddings find wide applications throughout the Pinterest search stack, primarily in retrieval and ranking tasks. Figure~\ref{fig:stack} presents a simplified depiction of the search retrieval and ranking stack at Pinterest and highlights the integration points for \modelname{} embeddings.

These embeddings are employed to power the retrieval of pins and products using HNSW~\cite{malkov2018efficient}. They are also instrumental in the L1 scoring model, where they enhance the efficiency of token-based retrieval sources. Moreover, \modelname{} embeddings serve as one of the most critical features in the L2 scoring and relevance models.

In this section, we delineate the results derived from the A/B tests we conducted. In these tests, production SearchSage embeddings were replaced with \modelname{} embeddings, resulting in boosted performance in both organic and promoted content (Ads) in search. 
Additionally, we provide results from a human relevance assessment conducted on actual production-sampled traffic. This evaluation further confirms the improved performance derived from the utilization of \modelname{} embeddings.
Finally, we demonstrate how employing query embeddings also enhances performance in other tasks, such as classification, particularly in situations where data availability is limited. This highlights the ability of the \modelname{} model to generalize to tasks different from its original training objectives.

\subsection{Human Relevance Evaluation}
To understand advantages of \modelname{}, we enlisted human evaluators to assess the relevance of candidates retrieved via two methods: \modelname{} embeddings-based pin retrieval and token-based pin retrieval. 

For this evaluation, we selected a set of $300$ queries, deliberately stratified across both head and tail queries. The top $8$ candidate pins were then retrieved from each system using these queries, and human evaluators determined the relevance of the pins to the corresponding query. Every query-pin pair received three judgements, with an inter-annotator agreement rate of $0.89$. Evaluation results revealed a noticeable improvement with \modelname{}, showing a $10\%$ increase in relevance compared to the token-based system.

\begin{figure}[t]
    \centering  
    \begin{subfigure}{0.45\linewidth}  
        \centering  
        \includegraphics[width=\textwidth, trim={20, 30, 395, 20}, clip]{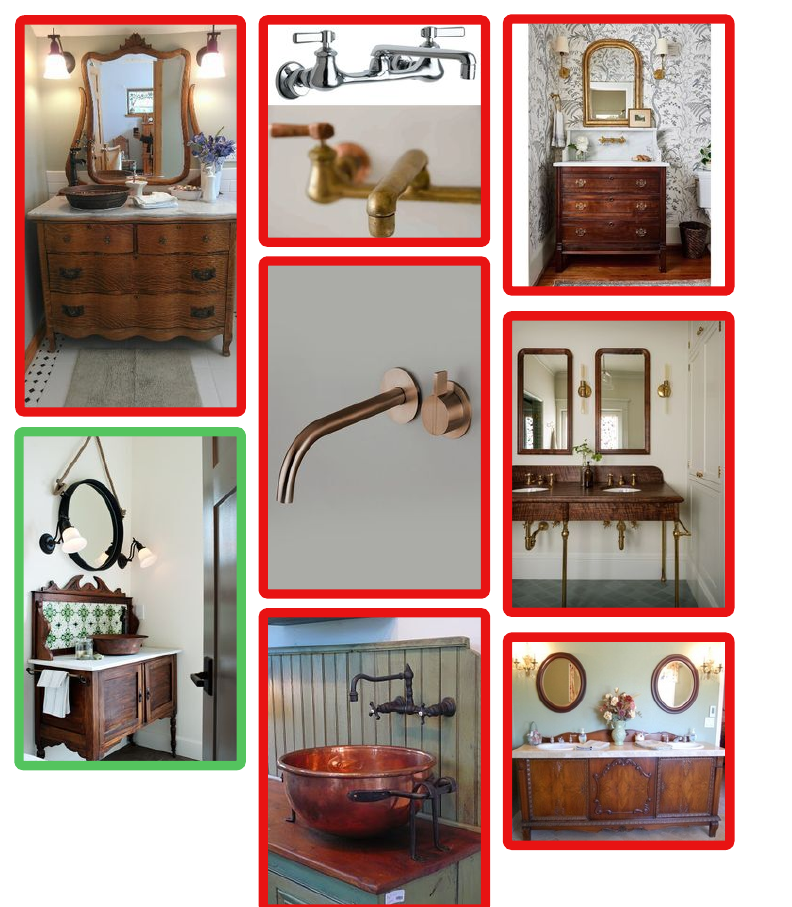}  
        \caption{Token-based}  
        \label{fig:figure1}  
    \end{subfigure}%  
    \begin{subfigure}{0.45\linewidth}  
        \centering  
        \includegraphics[width=\textwidth, trim={395, 30, 20, 20}, clip]{images/searchsage_paper_example_comparison.pdf}  
        \caption{\modelname{}-based}  
        \label{fig:figure2}  
    \end{subfigure} 
    \caption{Comparative display of pins retrieved in response to the query 'antique copper bathroom sink' from the token-based system and the \modelname-based system. Pins deemed relevant are outlined in green, while those considered irrelevant are encircled in red.}  
    \label{fig:example}  
\end{figure}

Figure~\ref{fig:example} offers a distinct comparison of retrieved pins for the query `antique copper bathroom sink' between the candidates retrieved by the token-based system and the \modelname-based system. The token-based retrieval system often fetches pins related to only part of the query and fails to fetch consistently relevant results. In striking contrast, nearly all pins retrieved by the \modelname-based system are highly relevant to the specified query, underlining the efficacy of the \modelname{} model in understanding the query and aligning similar pins and queries in the same space together.

\vspace{-7.5pt}
\subsection{Organic Search}
In this section, we outline the results of the A/B testing conducted to substitute the existing production SearchSage query and entity embeddings with \modelname{} embeddings for organic content within Pinterest search. Within the context of search experiments at Pinterest, our attention is largely concentrated on two key metrics: the search fulfillment rate and relevance. The search fulfillment rate is defined as the proportion of searches that result in a user engagement action of significance. Relevance is calculated as the weighted average relevance of the top eight pins for each query, assessed across different query segments. This is measured through human evaluation.

\begin{table}[htb]
\centering
\resizebox{\columnwidth}{!}{%
\begin{tabular}{@{}ccc@{}}
\toprule
                               & Search Fulfilment Rate     & Relevance                  \\ \midrule
Pin and Product Retrieval      & +4.1\%                      & +0.5\%                        \\
L1 Scoring                     & +0.5\%                      & +0.0\%                     \\
L2 Scoring and Relevance Model & +2.8\%                     & +3.0\%                        \\ \bottomrule
\end{tabular}%
}
\caption{Online A/B experiment results of \modelname{} in Organic Search.}
\label{tab:organic_results}
\end{table}

The impact on these two metrics, from replacing SearchSage with \modelname{}, is presented in Table~\ref{tab:organic_results}. The table provides data drawn from experiments for three distinct use-cases: (i) retrieval of pins and products, (ii)  L1 scoring model, and (iii) L2 scoring model and relevance model.

\subsection{Ads in Search}
The \modelname{} embeddings have also successfully replaced the SearchSage embeddings in various applications within Ads on Search surface. We present the results of three use cases: search engagement model, search relevance model, and product ads retrieval.

\begin{table}[tb]
\centering
\begin{tabular}{@{}ccc@{}}
\toprule
                               & gCTR         \\ \midrule
Product Ads Retrieval      & +5.27\%                                           \\
Ads Search Engagement Model                     & +2.96\%                                         \\
Ads Search Relevance Model & +1.55\%                                           \\ \bottomrule
\end{tabular}
\caption{Online A/B experiment results of \modelname{} for Ads in Search.}
\label{tab:ads_results}
\end{table}
Uniformly, we noted substantial improvements in engagement and relevance within Ads across all use cases. These increments, specifically in the long clickthrough rate (gCTR), are outlined in Table~\ref{tab:ads_results}. Furthermore, \modelname{} led to a noteworthy $4.95$\% increase in Ads relevance within the Search Ads relevance model. These gains highlight the positive impact of transitioning to \modelname{} embeddings for Ads on Search.

\vspace{-3pt}
\subsection{Classification}
One of the primary advantages of developing robust query representation such as \modelname{} is its utility in powering downstream applications, particularly when there is a lack of labels for learning large models. One example of this at Pinterest is interest classification, where we classify queries into a hierarchical taxonomy.
Using \modelname{} query embeddings for query representation, we were able to increase performance when compared to the baseline FastText~\cite{bojanowski2016enriching} model. Precision increased by $30\%$ on average across levels, with the larger gains coming from more granular levels. 

\section{Conclusion}
In this work, we presented \modelname, an end-to-end optimized set of query, pin, and product embeddings for Pinterest search, which have shown value across many applications.

In contrast to other work focused on learning embeddings for search, we demonstrate the value of unified query, pin, and product embeddings as both candidate generators and features in Pinterest search. We show a great improvement over previous solutions at Pinterest can be attributed to rich document text representations, which improved offline evaluation metrics by $>50\%$. We also describe practical decisions enabling serving and adoption, including compatibilty encoders, multi-task learning, and long-TTL caching.

Lastly, we summarize results from online A/B experiments across organic and ads applications, which have directly led to cumulative gains of $+7.4\%$ fulfilment rate on searches, and $+3.5\%$ relevance.

%%
%% The next two lines define the bibliography style to be used, and
%% the bibliography file.
\newpage
\bibliographystyle{ACM-Reference-Format}
\bibliography{omnisearchsage_main}

%%
%% If your work has an appendix, this is the place to put it.
% \newpage
% \appendix

\end{document}